\documentclass{emulateapj} 
\usepackage{psfig}
\usepackage{apjfonts}
\usepackage{mathrsfs}
\usepackage{graphicx}

\begin{document}

\title{Potassium: a new actor on the globular cluster chemical evolution stage. \\ 
The case of NGC~2808 \footnotemark[1]}
\footnotetext[1]{Based on data obtained at the ESO Very Large Telescope under the programs 
072.D-0507 and 091.D-0329.}

\author{Alessio Mucciarelli\altaffilmark{2},
Michele Bellazzini\altaffilmark{3},
Thibault Merle\altaffilmark{4},
Bertrand Plez\altaffilmark{5},
Emanuele Dalessandro\altaffilmark{2},
Rodrigo Ibata\altaffilmark{6}
}

\footnotetext[2]{Dipartimento di Fisica \&  Astronomia, Alma Mater Studiorum, Universit\`a di
Bologna, Viale Berti Pichat, 6/2 - 40127, Bologna, Italy},   
\footnotetext[3]{INAF--Osservatorio Astronomico di Bologna, via Ranzani 1, I--40127
Bologna, Italy}
\footnotetext[4]{Institute d'Astronomie et d'Astrophysique, Universit\'e Libre de Bruxelles, CP.226, Boulevard 
  du Triomphe, 1050, Brussels, Belgium}
\footnotetext[5]{Laboratoire Univers et Particules de Montpellier, Universit\'e Montepellier 2, CNRS, 
  F-34095 Montpellier, France}
\footnotetext[6]{Observatoire Astronomique, Universit\'e de Strasbourg, CNRS, 11, rue de l'Universit\'e, F-67000,
  Strasbourg, France}

\begin{abstract}

We derive [K/Fe] abundance ratios for 119 stars in the globular cluster NGC~2808,
all of them having O, Na, Mg and Al abundances homogeneously measured in previous works.
We detect an intrinsic star-to-star spread in the
Potassium abundance. Moreover [K/Fe] abundance ratios display statistically significant
correlations with [Na/Fe] and [Al/Fe], and anti-correlations with [O/Fe] and [Mg/Fe].
All the four Mg deficient stars ([Mg/Fe]$<$0.0) discovered so far in NGC~2808 
are enriched in K by $\sim$0.3 dex with respect to those with normal [Mg/Fe].
NGC~2808 is the second globular cluster, after NGC~2419, where a clear  
Mg-K anti-correlation is detected, albeit of weaker amplitude. 
The simultaneous correlation/anti-correlation of [K/Fe] with all  
the light elements usually involved in the chemical anomalies observed in globular cluster 
stars, strongly support the idea that these abundance patterns are due to the same 
self-enrichment mechanism that produces Na-O and Mg-Al anti-correlations. 
This finding suggests that detectable spreads in K abundances may be typical 
in the massive globular clusters where the self-enrichment processes are observed to
produce their most extreme manifestations.

\end{abstract}


\section{Introduction}

In the last decade, accurate photometric and spectroscopic investigations have changed our concept 
of globular clusters (GCs), which were considered for a long time as aggregates of coeval stars,  
born with the same initial chemical composition.
This traditional paradigm is still valid in terms of the 
iron and iron-peak elements abundance, while large variations in light elements (O, Na, Mg and Al)
were revealed in all the GCs studied so far, in our Galaxy \citep[see e.g.][]{carretta09,gratton12} 
and in the Local Group \citep{m09}. 

Chemical inhomogeneities are usually explained by invoking
the existence of (at least) a second generation of stars formed within the first $\sim 100$ Myr 
of the cluster life, from intra-cluster medium polluted by the first generation stars. 
Intermediate mass Asymptotic Giant Branch (AGB) stars \citep{dercole08},
Fast Rotating Massive Stars \citep[FRMS][]{decressin07} and binary stars \citep{demink09} were proposed 
as the most likely polluters, that should significantly affect the abundance of light elements 
(e.g. enhancing Na, Al and He and depleting Mg and O) while leaving the iron abundance unaffected.
However, all these models face serious problems and we are still far from a full
understanding of the processes underlying the presence of multiple 
populations in GCs \citep[see e.g. the alternative scenario proposed by][]{bastian13}.

Potassium is a {\sl new entry} among the chemical anomalies in GCs.
From the analysis of DEIMOS@Keck low-resolution spectra of 49 giant stars in the remote, massive GC NGC~2419, 
\citet[][Mu12 hereafter]{m12} discovered that its stars span an unusually large range in K abundances, 
from solar values up to [K/Fe]$\sim+2$ dex.
Also, [K/Fe] anti-correlate with [Mg/Fe] that also covers a huge range,
from values compatible with the $\alpha$-enhancement observed in other ancient GCs down 
to [Mg/Fe]$\sim-1$ dex. About 40\% of the stars in NGC~2419 
show sub-solar [Mg/Fe] abundance ratios and high K abundances. 
This were confirmed by \citet{cohen12} from the analysis of HIRES@Keck high-resolution spectra 
of 13 giant stars of NGC~2419. 
Note that most of the GC stars have [Mg/Fe] abundance ratios 
between +0.3 and +0.6 dex (thus, with variations generally smaller than those observed in the other light elements), 
and only a handful of GC stars with --0.3$<$[Mg/Fe]$<+0.3$ are known \citep{carretta09}.
NGC~2419 represents the only exception observed so far.

Currently, a solid interpretation of the origin of such a Mg-K anti-correlation is still lacking. 
\citet[][V12 hereafter]{ventura12} proposed a theoretical model where the Mg-poor/K-rich stellar population 
is {\sl an extreme population directly formed from the AGB and super-AGB ejecta}, 
supporting the first claim by Mu12 that the spread in K is ascribable to the same self-enrichment 
process able to produce the other chemical anomalies.
However, a proper modeling of the observed pattern requires a fine tuning in the reaction cross-sections 
and burning temperatures. 

%%%%OTHER CLUSTERS (CARRETTA+13)
\citet{carretta13} analysed small samples of stars in different evolutionary stages 
(turnoff, sub-giant and giant stars) in seven GCs, finding no evidence 
of intrinsic scatter in their K content. All the studied stars have Mg and K 
abundances that well match the values of the {\sl normal} stellar population 
of NGC~2419. 

%%%%PROJECT
In order to unveil the possible role played by K in the framework of the 
GC self-enrichment process, and to understand whether the Mg-K anti-correlation in NGC~2419 
is a singular event among the chemical anomalies of the GCs, we started a project aimed at deriving 
the K abundance in different GCs. 
This Letter is focused on the massive GC NGC~2808, one of the clusters with the most 
extended Na-O anti-correlation \citep[see e.g.][]{carretta06} and harboring 4 Mg-poor stars 
\citep{carretta09,carretta14}, making it an ideal cluster where K abundance spreads can be searched for.

\section{Observations}
The observations were performed with the spectroscopic facility 
FLAMES \citep{pasquini} in the UVES+MEDUSA combined mode.
We adopted the 860 Red Arm CD4 UVES set-up, with a spectral coverage 
of 6600-10600  \AA\ and a resolution of $\sim$45000, 
and the HR18 GIRAFFE grating, covering from 7648 to 7889 \AA\ 
and with a spectral resolution of 18400.
These set-ups were chosen in order to sample the K~I resonance line at 7699 \AA\ .  

We used two fibre allocation configurations to observe our targets, that are 119
Red Giant Branch (RGB) stars whose membership to NGC~2808 was already 
established, and whose Fe, O, Na, Mg, and Al abundances were homogeneously estimated by 
\citet[][C06 and C09 hereafter, respectively]{carretta06,carretta09}.
The goal of our project is to build on their analysis by adding the abundance of Potassium. 
In the following all the adopted abundance ratios are from these works, except for [K/Fe].

For each fibre configuration two exposures of 1300 sec each were secured, 
in order to reach a S/N ratio per pixel around the K line of $\sim$100 
for the faintest targets (V$\sim$15.4) and of $\sim$200 for the brightest ones (V$\sim$13.8).
All the spectra were reduced with the UVES-FLAMES and GIRAFFE ESO pipelines, including 
bias subtraction, flat-fielding, wavelength calibration, spectral extraction and 
order merging (only for UVES spectra). 

\section{K abundances}
The K abundances of the targets were derived from the measure 
of the K~I line at 7699 \AA\  with the package {\tt GALA} \citep{m13g}.
The line equivalent widths (EWs) were measured with the code {\tt DAOSPEC} \citep{stetson}, 
through the wrapper {\tt 4DAO} \citep{m13_4dao}. 
Eleven stars were observed both with UVES and GIRAFFE spectra: we found an average 
difference between their K line EWs of $EW_\mathrm{UVES}-EW_\mathrm{GIR}=2.5\pm1.2$ m\AA\  $(\sigma=4.0$  m\AA ), 
corresponding to a typical variation smaller than 2\% in EW ($\sim$0.03-0.04 dex in K abundance).

Effective temperatures ($T_\mathrm{eff}$) and surface gravities (log~g) are from C06. 
The analysis of the entire FLAMES sample of 4 GCs secured 
within this project (Mucciarelli et al. in prep.) has revealed 
that the use of the micro-turbulent velocities ($v_\mathrm{turb}$) derived by C06
leads to an unexpected trend between K abundances and $v_\mathrm{turb}$, 
probably due to the small number of Fe~I lines used to infer $v_\mathrm{turb}$.
In order to use a homogeneous scale of $v_\mathrm{turb}$, we adopted the calibration by \citet{kirby09} 
that provides $v_\mathrm{turb}$ as a function of log~g. No trend between $v_\mathrm{turb}$ and 
the K abundances are detected when the relation by \citet{kirby09} is used.
The K abundances are corrected for the departures from Local Thermodynamic Equilibrium (LTE) 
by applying non-LTE abundance corrections computed with the non-LTE radiative transfer 
code MULTI \citep[modified version 2.3,][]{carlsson86}, MARCS model atmospheres \citep{gustaf08} and a 
new model atom of neutral potassium (Merle et al. in prep.). 
NLTE EWs are larger than in LTE, leading to NLTE K abundances lower by about -0.3 dex.

Uncertainties in the derived abundances were computed by taking into account two different 
sources of errors: (i) the uncertainty arising from the EW measurement, as estimated by {\tt DAOSPEC}, 
typically of the order of $\sim$0.02-0.03 dex for GIRAFFE targets and $\sim$0.04-0.05 dex for UVES targets, and
(ii) the uncertainties arising from the atmospheric parameters. For the latter source of error, 
we take into account the correlation among the parameters: we adopted 40 K as typical uncertainty 
in $T_\mathrm{eff}$, as quoted by C06. A $T_\mathrm{eff}$ variation of $\pm$40 K changes 
the K abundance by $\pm$0.06-0.07 dex and  
leads to a variation of log~g smaller than $\pm$0.1 (with a null impact on the K abundance). This log~g 
variation leads to a change in $v_\mathrm{turb}$ of $\mp$0.04 km/s, corresponding to a K variation 
of $\pm$0.02-0.03 dex. Hence, we estimated a total uncertainty 
of about 0.09--0.11 dex.

\section{Results}

\begin{figure*}
\plottwo{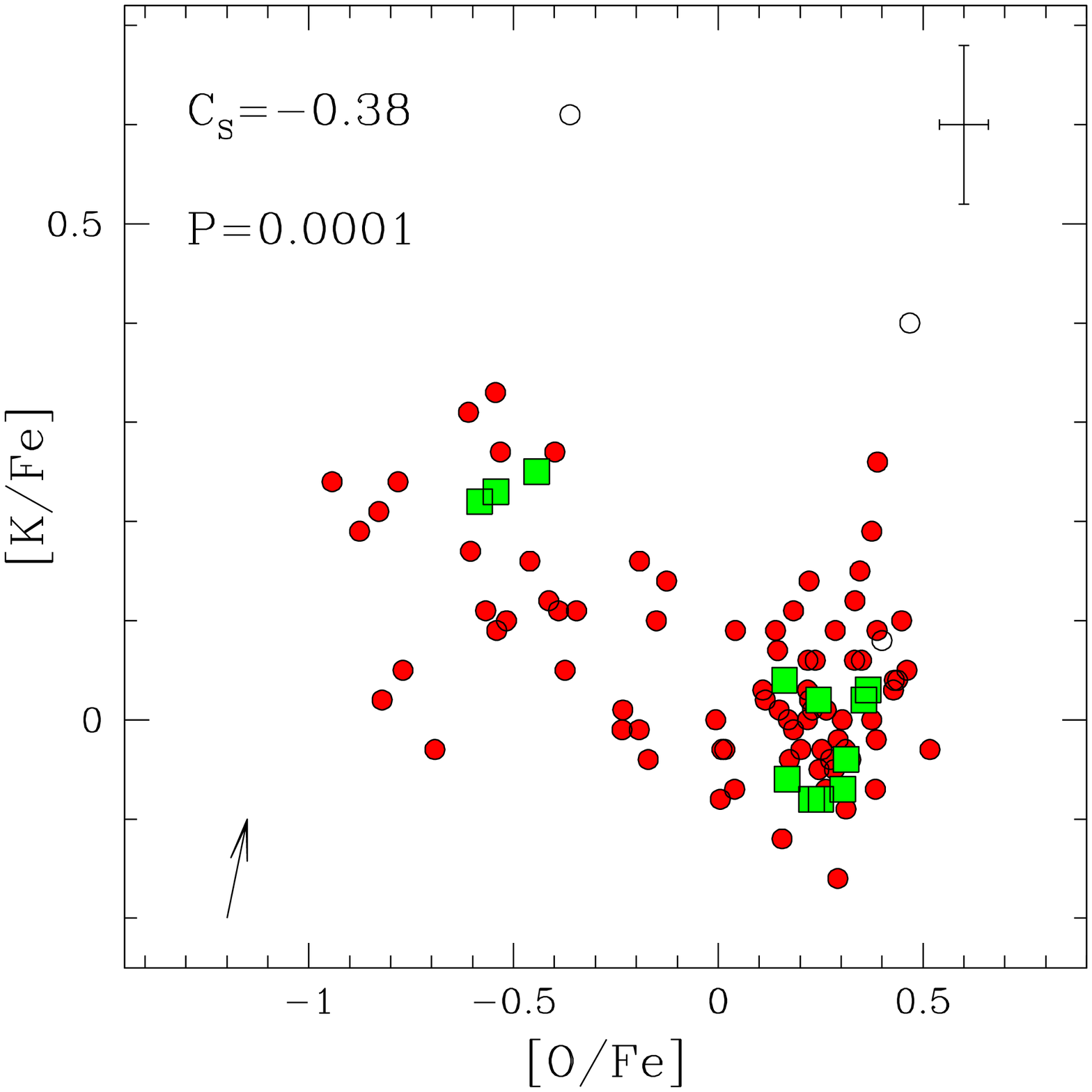}{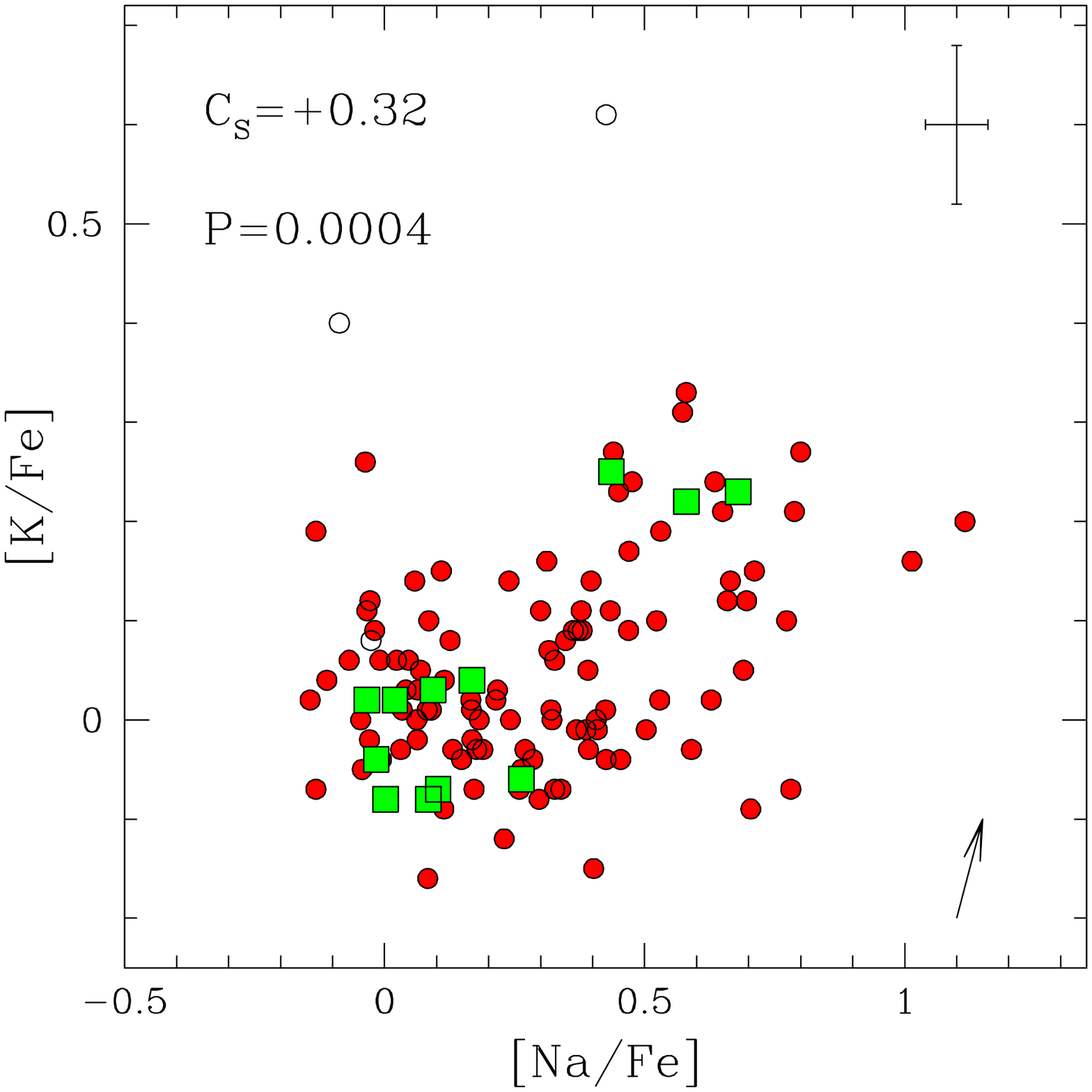}
\caption{[K/Fe] as a function of [O/Fe] (left panel) and [Na/Fe] (right panel). 
Red circles are the GIRAFFE targets and green squares the UVES targets. Empty circles are 
stars excluded from the analysis for their variable $V_r$ (likely binary stars). 
The arrows show the effects of a change by 40 K in $T_\mathrm{eff}$ and the corresponding 
variations in log~g and $v_\mathrm{turb}$.}
\label{kona}
\end{figure*}

The mean difference between our radial velocity estimates and those by C06 is $\langle \Delta V_r\rangle=-0.1$~km/s with 
a standard deviation of $\sigma_{\Delta}=0.8$~km/s. Three of the 119 targets have $\left|\Delta V_r\right|>3.0$~km/s, i.e. 
much larger than 3$\sigma_{\Delta}$, hence they were flagged as possible binary stars and excluded from the following 
discussion (but shown in the plots as empty symbols).
It is interesting to note that some of the candidate 
unresolved binary systems identified here
stand out as extreme outliers in the chemical abundance plots shown in Fig.~\ref{kona}. This suggests that
some stars observed to lie out of the general trends between abundance ratio in GCs or other stellar systems
may not have a genuine anomalous composition but, instead, may have their abundances 
altered by their unrecognised binary nature.
The abundance distribution of the remaining 116 stars spans a sizeable range, from [K/Fe]=--0.16 up to 
[K/Fe]=+0.33. We used the Maximum Likelihood algorithm described 
in Mu12, that provides mean and intrinsic spread of a given abundance ratio by taking into account 
the uncertainties of each individual star. We found that the sample displays a statistically significant 
{\em intrinsic} star-to-star scatter in [K/Fe], $\sigma_{int}^{[K/Fe]}$=0.05$\pm$0.01.

\begin{figure*}
\plottwo{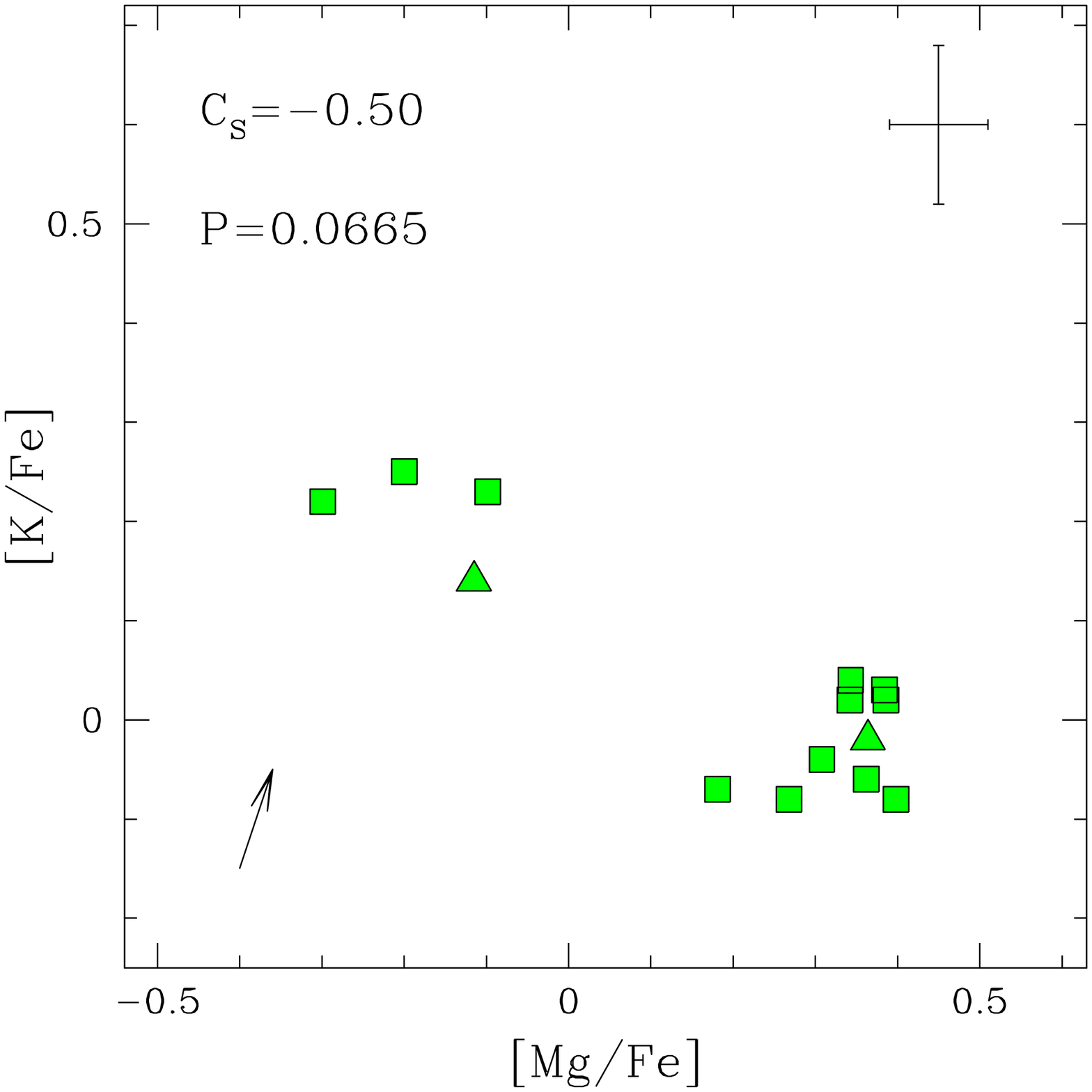}{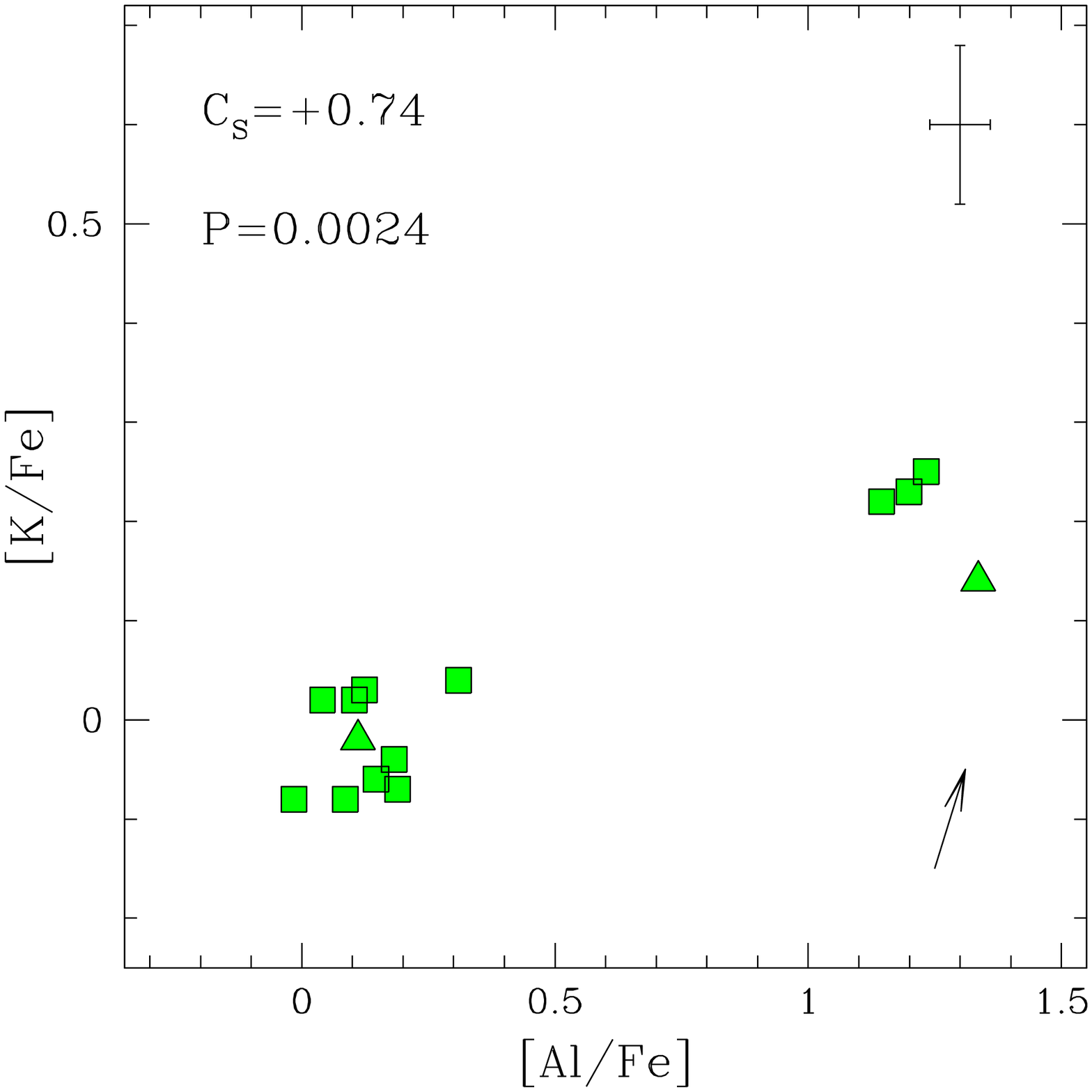}
\caption{[K/Fe] as a function of [Mg/Fe] (left panel) and [Al/Fe] (right panel). 
Green squares are the UVES targets, while the triangles are the additionally two stars in common 
with \citet{carretta14}.}
\label{kmgal}
\end{figure*}

Additional support for this conclusion comes from the behavior of [K/Fe] 
as a function of [O/Fe], [Na/Fe], [Mg/Fe] and [Al/Fe] shown in Fig.~\ref{kona} and Fig.~\ref{kmgal}.  
The GIRAFFE targets (for which only O and Na abundances are available) 
are shown as red circles and the UVES targets (for which O, Na, Mg and Al were measured) 
as green squares. 
Recently, \citet{carretta14} re-analysed the stars already discussed 
in C09 and derived new Mg and Al abundances for the stars 
originally discussed in \citet{carretta04} but not planned in our observations. 
However, two of these new stars are in common with our GIRAFFE sample and we included their Mg and Al 
abundances (marked as filled triangles in Fig.~\ref{kmgal}), even if formally their analysis 
is based on a slightly different $T_\mathrm{eff}$ scale (with $T_\mathrm{eff}$ differences of 71~K and --18~K 
with respect to C06).

All the four abundance ratios show clear and statistically significant correlations
with the K abundance. In particular, [Na/Fe] and [Al/Fe] correlate with [K/Fe], and 
[O/Fe] and [Mg/Fe] anti-correlate with [K/Fe]. 
Spearman correlation test gives probabilities that the variables are non-correlated of $P(K-O)=1.5\times 10^{-4}$,
$P(K-Na)=3.9\times 10^{-4}$, $P(K-Mg)=0.0665$, and $P(K-Al)=0.0024$.
To better assess the significance of the 
correlations of K with Mg and Al, we test the hypotheses that stars with [Mg/Fe]([Al/Fe])$\le 0.0(+0.5)$ and 
those with [Mg/Fe]([Al/Fe])$> 0.0(+0.5)$ ({\em a}) are extracted from the same parent distribution in [K/Fe], 
with a Kolmogorov-Smirnov test, and ({\em b}) 
are extracted from distribution of [K/Fe] having the same mean, with Student's and Welch's tests.
The hypothesis {\em a)} is rejected with $P=1.8\times10^{-3}$ and hypothesis {\em b)}
with $P<5.0\times10^{-5}$.  

At the core of the Mg-K anti-correlation is the fact that
the four Mg-deficient stars ([Mg/Fe]$< 0.0$) identified in NGC~2808, are also found to be K-enhanced 
with respect to Mg-normal stars.  While the amplitude
of the enhancement is relatively weak ($\simeq +0.3$~dex) the trend is analogous to that found in NGC~2419
by Mu12. Following the classification scheme set up by \citet{carretta14}, Mg-deficient and K-rich stars should
correspond to the {\em Extreme} population and Mg-normal and K-poor stars to the {\em Primordial} population.

\section{Sanity checks}

We performed several sanity checks to assess 
the robustness to systematics of the results described above.

%%% PARAM UNCERT
{\sl Correlations with atmospheric parameters} ---
Atmospheric parameters of stars along the RGB of a globular cluster are correlated.
An increase of $T_\mathrm{eff}$, coupled with the corresponding variations in log~g and $v_\mathrm{turb}$,
produces an increase in all the abundance ratios. The effect of these changes on the abundance 
ratios is shown by the arrows in Fig.~\ref{kona} and \ref{kmgal}. 
The directions of the plotted vectors show
that while a systematic in one of the atmospheric parameter can (in principle) play a role in the K-Na
correlation (and, perhaps, also in the K-Al one), they are almost perpendicular
to the directions of the K-Mg and K-O anti-correlations (see also the discussion in Mu12).

\begin{figure}
\plotone{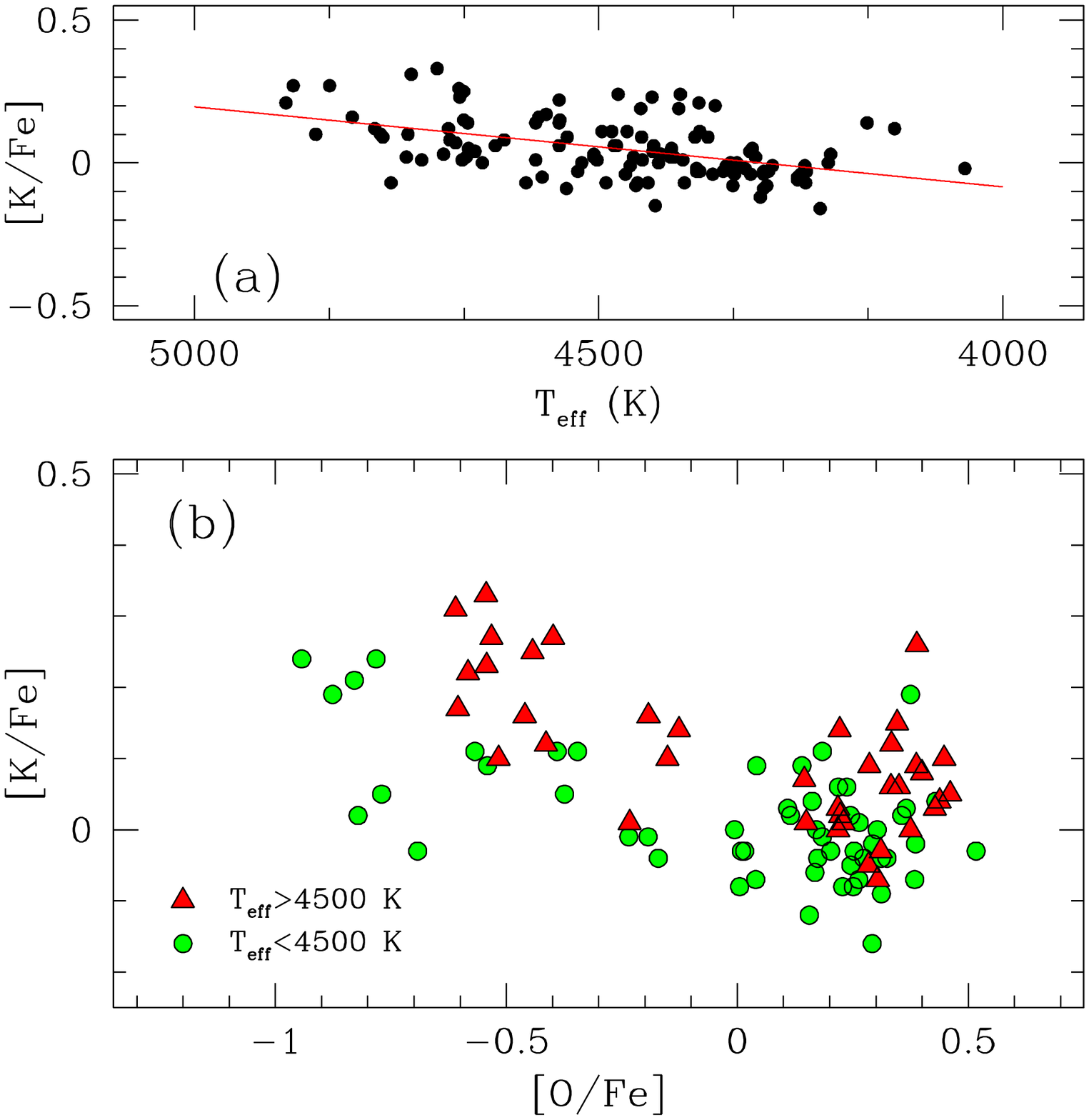}
\caption{Panel (a): trend of [K/Fe] abundance ratio with $T_\mathrm{eff}$. The red line is a linear fit to the data. 
Panel (b): [K/Fe] as a function of [O/Fe]. Stars hotter  and cooler  than 4500~K are plotted with 
different symbols (red triangles and green circles, respectively) to highlight the effect of the trend between K abundance and 
$T_\mathrm{eff}$ on the [K/Fe]-[O/Fe] correlation.}
\label{trend}
\end{figure}

Indeed, we noted that a weak residual trend between [K/Fe] and $T_\mathrm{eff}$ is present in our data 
(see panel {\sl (a)} of Fig.~\ref{trend}).
However, this trend with $T_\mathrm{eff}$ is clearly not at the origin of the K-O and K-Mg
correlation, as it can be appreciated from Fig.~\ref{trend}, where we plotted with 
different symbols stars hotter and cooler than $T_\mathrm{eff}$=~4500 K. Both groups display
the K-O correlation, with a small offset in K abundances fully consistent with the vector
shown in Fig.~\ref{kona}. The application of a rigid offset of +0.08 dex  in [K/Fe] to the stars 
of one group would lead to a full overlap of the two parallel sequences in the [K/Fe] vs.
[O/Fe] plane. Note that such an offset corresponds 
to a variation of $\sim$50~K, compatible with 1$\sigma$ uncertainty in $T_\mathrm{eff}$, 
but with a negligible impact on the abundances of other elements. A similar behavior occurs
in the [K/Fe] vs. [Mg/Fe] plane, though the small sample prevents an interpretation as clean as that emerging
from Fig.~\ref{trend}. 
The same effect is observed in the [K/Fe] vs. [Na/Fe] plane, explaining (at least partially) the 
large K abundance scatter measured at different values of [Na/Fe].
It is especially relevant to note that the removal of this
small systematic would made the observed correlations of K with other light elements even tighter
than that shown in Fig.~\ref{kona} and \ref{kmgal}.
Note that the three K-rich stars observed 
with UVES have a [K/Fe] scatter smaller than that measured among the GIRAFFE targets with similar [O/Fe] or [Na/Fe]. 
This can be explained with the fact that these stars have similar $T_{\rm eff}$, covering a range of about 100 K, while 
the entire sample of GIRAFFE targets covers about 800 K.

%NLTE
{\sl NLTE corrections} ---
[K/Fe] abundance ratios were re-derived using the NLTE corrections 
by \citet{takeda02}\footnote{http://optik2.mtk.nao.ac.jp/\~takeda/potassium\_nonlte/}, 
leading to a decrease of [K/Fe] of $\sim$0.2 dex.
The use of these NLTE corrections 
does not erase the observed correlations between [K/Fe] and the other light elements, and
the difference in K abundance between Mg-normal and Mg-deficient stars remains the same. 
The correlations are preserved also when a constant NLTE correction of --0.3 dex is adopted,
as done by Mu12 in the case of NGC~2419.

%SPECTROSCOPIC ANALYSIS
{\sl An alternative set of spectroscopic parameters} ---
We derived spectroscopically the atmospheric 
parameters of the stars of NGC~2808 observed with UVES, thus repeating the
analysis with a set of parameters fully independent of those by C09 that we used
before.
We combined the UVES spectra taken within our program, with those 
by C09 from the Red Arm 580 set-up, 
thus providing a total spectral coverage of about 4000 \AA\ . 
The atmospheric parameters were derived imposing 
{\sl (a)} 
no trend between Fe~I abundances and excitation potential (to constrain $T_\mathrm{eff}$), 
{\sl (b)} 
no trend between Fe~I abundances and EWs (to constrain $v_\mathrm{turb}$), 
{\sl (c)}
the same abundance, within the uncertainties, from Fe~I and Fe~II lines (to constrain log~g). 
This new analysis, based on $\sim$250 Fe~I lines and $\sim$20-25 Fe~II lines, 
provides atmospheric parameters that match well those used above, with differences smaller 
than 50-60 K in $T_\mathrm{eff}$, $\sim 0.1$ in log~g, and 0.2-0.3 km/s in $v_{t}$. 
Consequently, the newly derived abundances are in good agreement with those
presented above and the relevant trends with [K/Fe] remain essentially untouched.

\section{Discussion}

The present analysis provides two main results: 
(1)~the K abundance is not uniform among the stars of NGC~2808 but it exhibits a 
small but significant intrinsic spread; 
(2) K abundances are correlated with the abundances of the light elements involved 
in the chemical anomalies ubiquitously detected in globular clusters (namely O, Na, Mg and Al). 
In particular, the four Mg-deficient cluster stars detected so far 
are all K-enhanced, implying the existence of a tight Mg-K anti-correlation 
similar to that observed in NGC~2419, albeit with a much smaller amplitude. 
{\sl This is the second case of a GC where a Mg-K anti-correlation is detected.}

As for other light elements in NGC~2808 \citep{carretta14} the distribution of K
is not unimodal, thus suggesting that the enrichment occurred in discrete events.
The bi-modality of the K distribution is evident in the UVES sample. The GIRAFFE sample
satisfies all the criteria established by \citet{gmm} for a non-unimodal distribution 
(the so-called Gaussian mixture modeling, GMM, test).
In particular, the GMM test rejects the hypothesis of unimodal distribution at the 99.5\%
confidence level.

Stars with sub-solar [Mg/Fe] abundance ratios are quite rare in 
Galactic GCs: among the 19 GCs investigated by \citet{carretta09} 
only NGC~2808 is found to harbor four stars with [Mg/Fe]$<$0; two
were detected in M54 \citep{carretta10}, and four in Omega Centauri 
\citep{norris95}. All these clusters (plus NGC~2419) are among the most massive of the entire
GC system of the Milky Way and present the most extreme manifestations
of the light element self-enrichment process typical of globulars. They 
display extended (anti-)correlations between abundances of light elements,
complex Horizontal Branch morphologies with extended blue tails, and are observed
(or suspected) to host populations with extreme He abundances 
\citep[$Y>0.3$, see, e.g.][and references therein]{f2808,dalex08,grattomega,king,dalex11,jack}.
Our results for NGC~2419 and NGC~2808 strongly suggest that, at least in these
massive and metal-poor clusters, Mg-deficiency is associated with K-enhancement.
Other clusters, where self-enrichment did not reach the stage leading to 
the production of Mg-deficient stars, probably were not able to synthesise enough Potassium
to display a detectable spread in [K/Fe] \citep{carretta13}. 

The evidence that the [K/Fe] correlates also with O, Na, and Al supports the hypothesis 
that the spread in K is ascribable to the same self-enrichment 
process that produces the observed abundance variations (and correlations) for these elements. 
Still, placing NGC~2419 in this scenario is not straightforward.
In fact, while its K-Mg anti-correlation is consistent with being an extended version of
that observed in NGC~2808, this is not the case for the K-Na and K-Al correlations, since, 
according to \citet{cohen12}, at any given [Na/Fe] and [Al/Fe] there is a large spread of [K/Fe] 
\citep[see also][for a thorough discussion]{carretta13}.
Perhaps this is somehow related to the different metallicity regime of the two clusters, whose
[K/H] ratio for first-generation stars differ by $\sim 1.0$~dex (see also V12).

Within the scenario where the self-enrichment is driven by AGB stars, 
K production is expected to occur in the same 
hot bottom burning environment where the other chemical anomalies 
are produced (V12). In these models, the stars of the extreme populations of both NGC~2419 and NGC~2808
are thought to have formed directly from the pure ejecta of AGB/super-AGB stars, 
without dilution with pristine gas \citep[see e.g][]{f2808,dicriscienzo11}.
Potassium can be produced by proton capture on Argon nuclei, through the 
thermonuclear chain $^{36}Ar(p,\gamma)^{37}K(e^{+},\nu)^{37}Cl(p,\gamma)^{38}Ar(p,\gamma)^{39}K$, 
occurring at temperatures of about $10^8$ K. 
Even if the chain able to produce K is known, 
several uncertainties affect our knowledge of this reaction. 
In particular, V12 had to increase the cross section of the 
reaction $^{38}Ar(p,\gamma)^{39}K$ by a factor of 100 to reproduce the K abundances 
measured in the Mg-deficient stars of NGC~2419. We refer the reader to V12 for a complete 
discussion about the uncertainties in these reactions.

Note that no explicit predictions about K abundance anomalies are available for 
other models of self-enrichment (based on FRMS or binary stars) 
and we cannot use the observed scatter in the K content of NGC~2808 to 
discriminate among different scenarios. 
However, strong Mg-depletion can be currently predicted only by AGB/super-AGB models, 
with adequate mass loss rates and/or convection efficiency. It is also interesting to
note that V12, in their basic model for a cluster as metal-poor as NGC~2419, predict that 
Mg-depleted stars should also have their oxygen abundance depleted by as much as $\simeq -1.7$~dex,
not much different from the maximum oxygen depletion we observe in NGC~2808 stars ($\simeq -1.3$~dex).

In conclusion, the present analysis supports the idea
that K enrichment is among the outcomes of the process of self-enrichment occurring in globular
clusters and, in particular, significant K-enhancement is present in extremely Mg-depleted
stars. Hence, at least for the Mg-K anti-correlation, 
NGC~2419 is not a {\em unicum}.
The future extension of this survey to other clusters will establish if 
the homogeneous analysis of large samples may reveal the presence of 
small spreads of K abundances also in other GCs.

\acknowledgments
We warmly thank the anonymous referee for suggestions in improving the paper.
M.B. acknowledges the financial support from PRIN MIUR 2010-2011 project ``The Chemical and 
Dynamical Evolution of the Milky Way and Local Group Galaxies'', prot. 2010LY5N2T.
T.M. is supported by the FNRS-F.R.S. as temporary post-doctoral researcher under grant No. 2.4513.11.

\end{document}